\documentclass[twocolumn,showpacs,amsmath,amssymb]{revtex4}

\usepackage{graphicx}
\usepackage{dcolumn}
\usepackage{bm}

\newcommand{\bs}{\boldsymbol}

\begin{document}

\preprint{APS/123-QED}

\title{Lagrangian statistics in forced two-dimensional turbulence}

\author{O. Kamps, R. Friedrich}

\affiliation{Institute of Theoretical Physics, University of
M\"unster, Wilhelm-Klemm-Str. 9, 48149 M\"unster, Germany}

\date{\today}

\begin{abstract}

We report on simulations of two-dimensional turbulence in the
inverse energy cascade regime.
Focusing on the statistics of Lagrangian tracer particles, scaling
behavior of the probability density functions of velocity
fluctuations is investigated. The results are compared to the
three-dimensional case. In particular an analysis in terms
of compensated cumulants reveals the transition from a strong
non-Gaussian behavior with large tails to Gaussianity. The reported computation of correlation functions for the acceleration components sheds light
on the underlying dynamics of the tracer particles.

\end{abstract}



\maketitle

\begin{figure}[t]
\includegraphics[width=.4\textwidth]{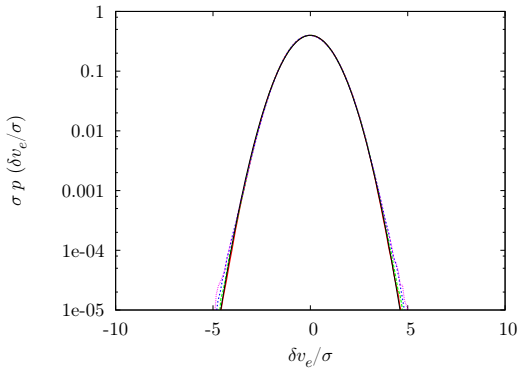}
\caption{Rescaled pdfs of the Eulerian velocity increments for the distances $ r = 0.08,0.1,0.2,0.6$. For comparison a Gaussian pdf is shown. Small deviations from Gaussianity exist and can be quantified by the moments \cite{boffetta00pre}. } \label{fig:pdfEuler}
\end{figure}

\begin{figure}[t]
\includegraphics[width=.4\textwidth]{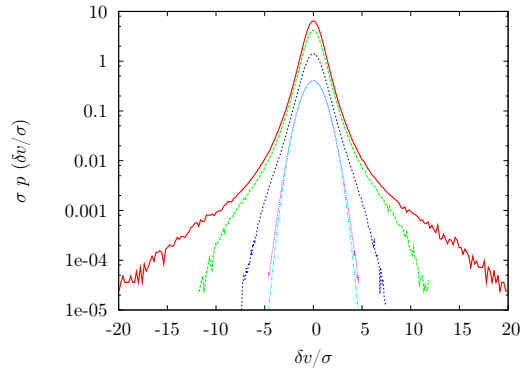}
\caption{Rescaled pdfs of Lagrangian velocity increments for the time
lags $\tau = 0.045,0.09,0.22,1.79 T_I$ (from outer to inner
curves). The most inner curve is a Gaussian pdf. The pdfs are vertically shifted.  } \label{fig:pdfGauss}
\end{figure}

\begin{figure}[b]
\includegraphics[width=.4\textwidth]{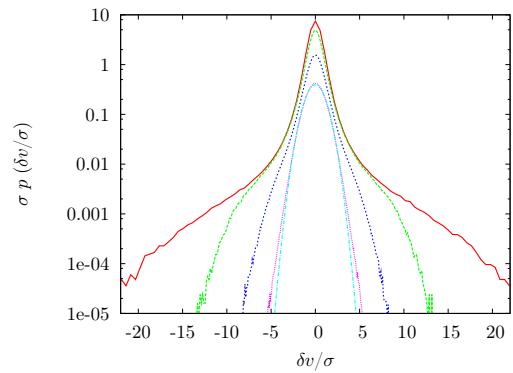}
\caption{Rescaled pdfs of Lagrangian velocity increments for the
same time lags as in Fig.\ref{fig:pdfGauss}. In this case the
forcing was confined to a small number of Fourier modes.}
\label{pdfDelta}
\end{figure}

\textit{Introduction} In recent years, the Lagrangian description of
turbulent  flows has attracted much interest from the experimental
point of view \cite{laporta01nat,mordant01prl} as well as in
numerical \cite{biferale05pof,homann06unp} and analytical
investigations \cite{friedrich03prl}. This is not only due to the
relevance of the Lagrangian approach for applications like turbulent
mixing and the dispersion of pollutants. But fundamental turbulence
research also benefits from this alternative description and its
relation to the Eulerian formulation. The turbulent state, for
example, is completely represented by the acceleration of the tracer
particles. The classical Kolmogorov theory (K41) predicts
self-similar scaling of the probability density functions (pdfs) of
the velocity increments in homogeneous isotropic turbulent flows. It
is a well known fact that in three dimensions the assumption of
self-similarity is violated for Eulerian velocity increments. This
is referred to as Eulerian intermittency. We know from recent
experiments and numerical simulations, that intermittency is
observed in the Lagrangian picture as well.
For two-dimensional turbulence the situation is different: although numerical
investigations show deviations from Gaussianity
\cite{boffetta00pre} for the Eulerian velocity increments, the
inverse cascade provides an Eulerian flow field with scaling
properties compatible with the K41 predictions. For this reason, the
inverse energy cascade in two-dimensional turbulence seems to be the
ideal system to shed light on the Lagrangian statistics and its
relation to the Eulerian case. \\
The paper in hand presents results from numerical investigations of the Lagrangian dynamics of forced two-dimensional turbulence. Thereby we focus on intermittency.
The remainder of this article is structured as follows. After briefly discussing the system under consideration and summarizing some facts regarding the numerics, we present results on the Eulerian increment statistics.
The main part of the article is devoted to the study of the Lagrangian increment pdfs. Finally we discuss properties of the acceleration
correlation.\\[1ex]
\begin{figure}[b]
\includegraphics[width=.4\textwidth]{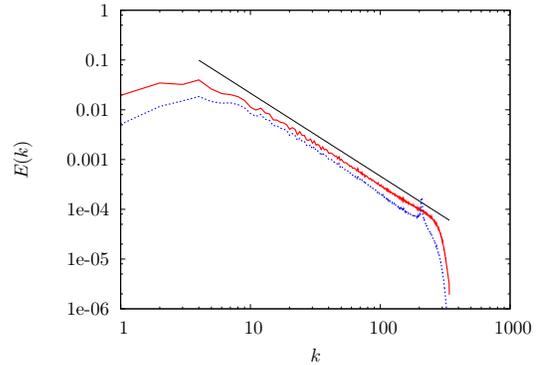}
\caption{Energy spectra $E(k)$ of simulations using the forcing with
the short (upper curve) and the long (lower curve) correlation
length. The line denotes the Kolmogorov prediction $E(k)\sim
k^{-5/3}$.} \label{fig:spectrum}
\end{figure}
\textit{Basic equation} The behavior of two-dimensional fluid
motion is governed by the vortex transport equation
\begin{equation}
\label{vortexTransportEquation}
\partial_ t \omega = -\bs{u} \cdot \nabla \omega + \nu \Delta \omega - \gamma \omega  -\Delta f ,
\end{equation}
with the vorticity $\omega = \omega({\bs x},t)$ and the velocity
$\bs{u}=\bs{u}(\bs{x},t)$. The velocity components ${ u_{x_1}} =
\partial_{x_2} \phi$ and ${ u_{x_2}} = -\partial_{x_1} \phi$ are
connected to the vorticity field $\omega$ via the stream function
$\phi$ by $\Delta \phi = -\omega$. We investigate two different
types of forcing $f$. One with a rapidly decaying spatial
correlation function $ \langle  f( {\bs x } + {\bs r }) f({\bs x })
\rangle \sim \exp (-r^2/2l_c^2) $ \cite{boffetta00pre} where $l_c$
is the length-scale of the correlation. The second is confined to a
small shell of wavenumbers in Fourier space \cite{frisch84pof}
leading to a long spatial correlation. Both forcings are
$\delta$-correlated in time and have the property to inject energy
at small scales into the system. The damping term $-\gamma \omega$
in (\ref{vortexTransportEquation})
extracts energy at large scales from the system and  avoids the generation of a large scale flow. \\[1ex]
%
\textit{Numerical method} The integration of
(\ref{vortexTransportEquation}) was performed by  a fully dealiased
pseudo-spectral method on a doubly periodic square domain with side
length $2\pi$ and $1024^2$ grid points. For numerical reasons, the
viscous term is replaced by a hyperviscous term of order eight. The
time evolution for the field and the tracer particles was achieved by a
fourth order Runge-Kutta scheme. After reaching stationarity we
introduced $10^5$ particles with random but equally distributed
initial positions into the flow and monitored their trajectories for
about 200 Lagrangian integral times $T_I$. The velocity and the
acceleration of the tracers were determined by a bicubic
interpolation scheme. Except for Fig.\ref{fig:spectrum} and Fig.\ref{pdfDelta} all
presented results are obtained using the forcing with the rapidly decaying spatial correlation.\\[1ex]
%
\textit{Eulerian statistics} In order to check the parameters for
the numerical integration, we first analyze the Eulerian velocity
field. Fig.\ref{fig:spectrum} shows the energy spectrum for both
kinds of forcing together with a line showing the K41-scaling
$k^{-5/3}$. For a stationary velocity field, the longitudinal
Eulerian velocity increments are defined as $\delta
v_e(\bs{x},\bs{r}) = \delta \boldsymbol{v}_e(\bs{x},\bs{r}) \cdot
\hat{\boldsymbol r}$ with $\delta \boldsymbol{v}_e(\bs{x},\bs{r}) =
{\boldsymbol u} ( \bs{x} + \bs{r} )- {\boldsymbol u}(\bs{x})$ and
$\hat{\boldsymbol r} = {\bs r}/r$. If we additionally assume
isotropy and homogeneity of the flow, we can write $\delta
v_e(\bs{x},\bs{r}) = \delta v_e(r)$. The pdfs $p(\delta v_e(r))$ are scaled to unit standart deviation by $\sigma p(\delta v_e(r) / \sigma )$ with $\sigma=\langle \delta v_e (r) \rangle^{1/2}$. Fig.\ref{fig:pdfEuler} shows
the rescaled pdfs for different $r$. The shape does  not vary with $r$ and hence the pdfs are self-similar.
This is in agreement with experimental \cite{paret98pof} and
numerical \cite{boffetta00pre} studies and leads to the conclusion
that intermittency is absent in the inverse
energy cascade as far as the Eulerian increments are concerned. \\[1ex]
\textit{Lagrangian velocity statistics} In the Lagrangian frame of
reference the velocities are recorded along the trajectories of
tracer particles $\bs{v}(\bs{y},t) = \left[ \bs{u}(\bs{x},t)
\right]_{\bs{x}=\bs{X}(\bs{y},t)}$ , where $\bs{y}$ is the starting
position of the tracer and $\bs{X}(\bs{y},t)$ is its current position. Velocity fluctuations are characterized by
the pdfs of the Lagrangian velocity increments $\delta v_i (\tau) =
v_i (t+\tau) - v_i(t)$ with $i=x_1,x_2$. Due to isotropy we average with respect to both spatial components. Therefore in the following the Lagrangian velocity increment will be denoted as
$\delta v (\tau)$.  The moments of $ p(\delta v (\tau)) $ are known
as the structure functions $S_p (\tau)= \langle \delta v (\tau)^p
\rangle$. Fig.\ref{fig:pdfGauss} shows pdfs of the Lagrangian
velocity increments for several time lags $\tau$. For time lags of
the order of $T_I$, the pdfs are close to a Gaussian distribution
whereas for small $\tau$ the pdfs show large tails.
%
\begin{figure}[b]
\includegraphics[width=.4\textwidth]{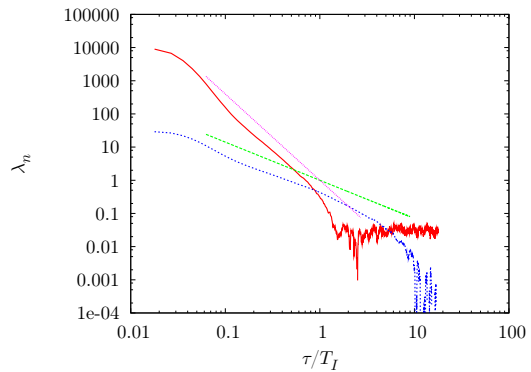}
\label{fig:cumulants} \caption{Compensated cumulants of order four
(lower curve) and six (upper curve) for 2D turbulence. As a guide for the eye $\tau^{-1.15}$ and $\tau^{-2.6}$ are shown. } \label{fig:cumulants}
\end{figure}
A rescaling of the distribution functions resulting in a collapse to a universal
distribution is not possible. Accordingly, classical Kolmogorov
scaling can not be observed in the Lagrangian frame in contrast to
the Eulerian case. Deviations from the Gaussian shape can be
quantified by the compensated cumulants $ \lambda_n  =
c_n/\sigma^n$. The $c_n$ are the cumulants connected to the
characteristic function $\hat{C}(k)$ of a pdf by
\begin{equation}
\hat{C}(k) = \exp \left[ \sum_{n}^{\infty} \frac{c_n}{n!}(ik)^n
\right]
\end{equation}
and $\sigma^2$ is the variance (corresponding to $c_2$).

\begin{figure}[t]
\includegraphics[width=.4\textwidth]{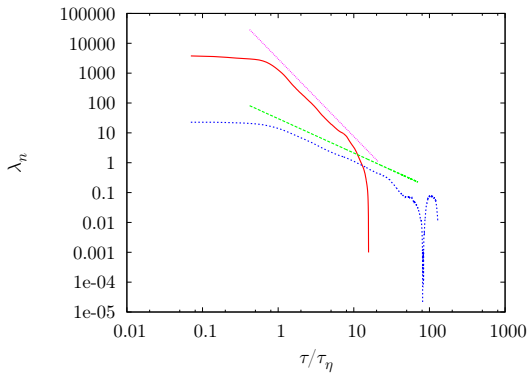}
 \caption{Compensated cumulants of order four
(lower curve) and six (upper curve) for 3D turbulence together with $\tau^{-1.15}$ and $\tau^{-2.6}$. The time is given in multiples of the Kolmogorov time $\tau_{\eta}$. }
\label{fig:cumulants3D}
\end{figure}

For a Gaussian distribution all $\lambda_n$ with an order $n$ higher
than 2 vanish. The compensated cumulants can easily be computed from
the structure functions $S_p$. For symmetric pdfs, $\lambda_4 =
S_4/S_2^2-3$ is the \textit{excess kurtosis} and the sixth order
normalized cumulant
reads $\lambda_6= (S_6 - 15 S_4 S_2)/S_2^3+30$.\\
In Fig.\ref{fig:cumulants} we see a log-log plot of $\lambda_4$ and
$\lambda_6$. The crucial point is that for a self-similar signal the
kurtosis should be a constant at least in the region where
self-similarity of the pdfs is expected. In our simulations the
kurtosis depends strongly on $\tau$ and for intermediate times
follows a power-law as can be seen in Fig.\ref{fig:cumulants}. This
also holds for $\lambda_6$ which however decays faster than
$\lambda_4$. As a reference we plotted two lines following the power
laws $\tau^{-1.15}$ and $\tau^{-2.6}$. During the decay of the
cumulants the pdfs converge to the Gaussian shape. In order to
investigate the universality of the observed behavior we used the
data provided by \cite{biferale05pof,biferale04prl} to calculate the
same quantities for three-dimensional turbulence (see
Fig.\ref{fig:cumulants3D}). Again scaling behavior can be detected
for intermediate times and for comparison we added power-laws with
the same exponents as in Fig.\ref{fig:cumulants}. For very small and
very large $\tau$ the shape of the compensated cumulants show
differences between 2D and 3D. We devote this to the fact that in 2D
the energy is injected on the small scales and mainly dissipated at
the large scales whereas in 3D the situation is complementary. Again
our results give strong evidence for intermittency in
two-dimensional Lagrangian turbulence.

\begin{table}[!h]
  \centering
  \begin{ruledtabular}
  \begin{tabular}{c|cccc}
      $p$  & 1 & 3 & 4 & 5   \\ \hline
      $\zeta^a_p$  & 0.557 $\pm$ 0.002 &  1.267 $\pm$ 0.007 & 1.35 $\pm$ 0.018 & 1.313 $\pm$ 0.033 \\
      $\zeta^b_p$  & 0.557 $\pm$ 0.003 &  1.313 $\pm$ 0.008 & 1.45 $\pm$ 0.019 & 1.588 $\pm$ 0.029 \\
  \end{tabular}
  \end{ruledtabular}
  \caption{ESS scaling exponents for 2D turbulence}
  \label{essTable}
\end{table}
%
\textit{Scaling of the Lagrangian structure functions} Additionally
to the cumulants, the structure functions are computed to
characterize the pdfs. As no scaling region is visible for the structure functions we have to
rely on the Extended Self Similarity (ESS) technique
\cite{benzi93pre} in order to estimate scaling exponents.
\begin{figure}[t]
\includegraphics[width=.4\textwidth]{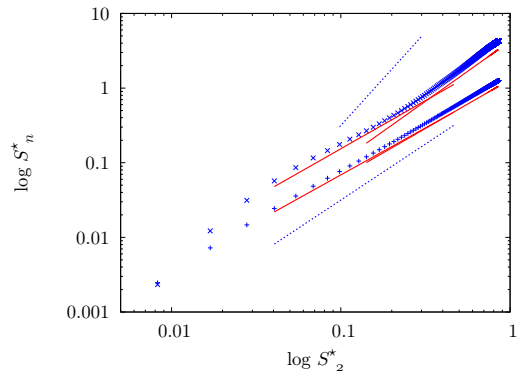}
\caption{ESS plot of $S^*_3$ (lower curve) and $S^*_5$ (upper
curve). The lines show the ESS scaling laws with exponents
$\zeta^a_p$ (lines ranging from $S^*_2=0.04$ to $S^*_2=0.4 $) and
$\zeta^b_p$ (from $S^*_2=0.015$ to $S^*_2=0.9 $) obtained by the two
fitting procedures (see text). The dashed lines correspond to the
K41 scaling.} \label{ess}
\end{figure}
%
To apply ESS we have to use the structure functions for the absolute
values of the velocity increments $ S^*_p (\tau)= \langle
|\delta v (\tau) |^p \rangle$. Standard arguments of dimensional
analysis lead to the scaling behavior $S^*_p \sim
\tau^{\zeta_p}$ with $\zeta_p = p/2$. The ESS-plot is shown in Fig.\ref{ess}.
Estimating the exponents up to order five in the spirit of \cite{homann06unp} leads to the
values $\zeta^a_p$ shown in Tab.\ref{essTable}. We also performed the analysis
for larger values of $S^*_2$ (corresponding to larger $\tau$) \cite{biferale05pof} resulting in the exponents $\zeta^b_p$. In both cases the exponents deviate strongly from the K41 predictions which is in agreement with the observation that the excess kurtosis is not constant. \\[1ex]
\textit{Dependence on the forcing} To study the effect of the
different forcings on Lagrangian observables, we also performed
simulations with a forcing limited to a small number of Fourier
modes. In this case, the results for the pdfs (Fig.\ref{pdfDelta})
as well as for the cumulants are qualitatively and quantitatively
similar to the situation with short spatial correlation. This leads
to the conclusion that the observed deviation of the Lagrangian pdfs
from the K41 prediction is very robust and seems to
be independent of the type of forcing. \\[1ex]
\textit{Acceleration correlations} The path of a Lagrangian tracer particle starting at the position $\bs{y}$ is
uniquely defined by the acceleration acting on the particle. The acceleration 
is given by the right-hand side of the Navier-Stokes equation
\begin{align}
\bs{a}&(\bs{y},t)  \nonumber \\
&= \left[ -\nabla p(\bs{x},t) + \nu \Delta \bs{u}(\bs{x},t) + \nabla
\times f(\bs{x},t) \right]_{\bs{x}=\bs{X}(\bs{y},t)}, \label{eq:acceleration}
\end{align}
where the pressure $p(\bs{x},t)$ is related to the vorticity by
$\Delta p = \nabla \cdot [\bs{u} \times \bs{\omega}] -
\frac{1}{2}\Delta \bs{u}^2$. Besides this fact, the acceleration is
also of central interest for turbulence modeling
\cite{obukhov5906aig,aringazin04pre}. Particularly in two dimensions,
it is convenient to split up $\bs{a}(\bs{x},t)$ into a component
parallel and a component perpendicular to the current velocity of the tracer. The latter is sensitive to circular
motions. The two components are defined as $a_{\|}=\bs{a}\cdot
\bs{u}/|\bs{u}|$ and $a_{\perp}=\bs{a}\cdot (-u_y,u_x)/|\bs{u}|$.
\begin{figure}[t]
\includegraphics[width=.4\textwidth]{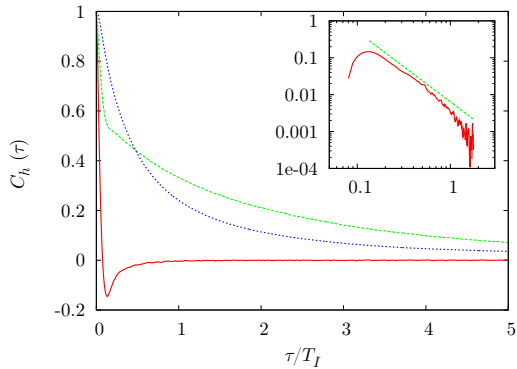}
\label{correlation} \caption{Correlation functions  $
C_{a_{\perp}}(\tau)$ (upper curve), $ C_{a_{\parallel}}(\tau)$
(lower curve) and $ C_{\bs{v}}(\tau)$ (middle curve). The inset
presents a log-log plot of $-C_{a_{\parallel}}(\tau)$. The line
corresponds to $ \tau^{-1.9}$. } \label{fig:correlation}
\end{figure}
As pointed out in \cite{mordant02prl}, long correlation times of the
velocity increments and the acceleration play a key role for the
occurrence of Lagrangian intermittency. This poses the question
whether the observed deviation from self-similarity of the increment
pdfs in 2D is also connected to long correlation times of the
acceleration. In Fig.\ref{fig:correlation} we can see the
correlation functions $C_{h}(\tau)$ with $h=a_{\perp},a_{\parallel},\bs{v}$ for the two components of the acceleration and for the velocity defined as $ C_{h}(\tau) = \langle h(t) h(t+\tau) \rangle / \langle h^2(t)
\rangle.$ The correlation for the parallel component decays very fast and
approaches zero after passing a minimum at negative values. The
perpendicular acceleration component also decorrelates very fast for small $\tau$. For $\tau$ bigger than the time corresponding to the
minimum of $C_{a_{\parallel}}(\tau)$ it bends off into a region where it decays
much slower leading to a very long correlation time. The same behavior
is also observed for 3D turbulence \cite{mordant04njp,laporta01nat,toschi05jot}. In \cite{toschi05jot}
it was related to the spiraling motion in a vortex filament. Here
we also observe events where the particle runs through loops.
Because the motion is confined to a plane, there is no movement in
the third spacial dimension that could contribute to the decorrelation
of $a_{\perp}$. For comparison also $C_{\bs{v}}$ is shown in Fig.\ref{fig:correlation}.
The inset in the same figure demonstrates that $C_{a_{\parallel}}$ follows a power-law in the range between
its minimum and $T_I$. The results for the temporal correlations of the acceleration suggest that the stochastic
process for the velocity increments is essentially non-Markovian, as has been emphasized in \cite{friedrich06prl}. \\[1ex]
\textit{Conclusion} We presented a detailed investigation regarding
the statistics of tracer particles in the inverse energy cascade
regime of two-dimensional turbulence. For different types of
forcing, we detect the same deviations from self-similarity for the
Lagrangian velocity increment pdfs. This is strong evidence in favor
of Lagrangian intermittency in the inverse energy cascade. It is of
particular interest as for the Eulerian frame no intermittency can
be detected. Any attempt to relate the two frames of reference has
to incorporate this fact. The observation that in 2D and 3D the
compensated cumulants show the same scaling behavior for
intermediate time lags suggest that the underlying dynamical process
exhibits a certain degree of universality independent of the
dimension. This view is supported by the fact that the acceleration
components show long time correlations which are similar to the 3D
case. The explanation for the scaling of the compensated cumulants
remains an open question.
\\[1ex]
\textit{Acknowledgements} We are grateful to H. Homann, R. Grauer
and M. Wilczek for fruitful discussions and acknowledge
support from the Deutsche Forschungsgesellschaft (FR 1003/8-1). We also thank the supercomputing center Cineca (Bologna, Italy) for providing and hosting of the data for 3D Turbulence.


\bibliographystyle{alphadin} 

\bibliographystyle{plain}           



\end{document}